\documentstyle[epsfig,12pt]{article}
\def\beqn{\begin{eqnarray}}
\def\eeqn{\end{eqnarray}}

\def\beq{\begin{equation}}
\def\eeq{\end{equation}}
\def\ba{\beq\new\begin{array}{c}}
\def\ea{\end{array}\eeq}

\begin{document}

\setcounter{footnote}0
\renewcommand{\thefootnote}{\fnsymbol{footnote}}

\vfill

\begin{titlepage}
\begin{center}
{\Large \bf William I. Fine Theoretical Physics Institute \\
University of Minnesota \\}
\end{center}
\vspace{0.15in}
\begin{flushright}
FTPI-MINN-10/16\\
UMN-TH-2908/10\\
ITEP-TH-23/10\\
June 2010
\end{flushright}

\begin{center}

{ \Large \bf   Remarks on Decay of Defects with Internal Degrees of Freedom}
\end{center}

\begin{center}
 { \bf A.~Gorsky\, and \bf M.~B.~Voloshin}
\end {center}
\vspace{0.3cm}
\begin{center}
{\it  William I. Fine Theoretical Physics Institute,
University of Minnesota,
Minneapolis, MN 55455, USA}\\
and \\
{\it Institute of Theoretical and Experimental Physics, Moscow
117259, Russia}
\end{center}

\vspace*{.45cm}
\begin{center}
{\large\bf Abstract}
\end{center}
We consider the decay of metastable walls and strings populated by
additional degrees of freedom.
The examples involve the decay of an axion wall in an external magnetic field,
pionic walls, metastable walls in dense QCD. It is
shown that the induced fermions escape from the
wall during the decay process providing an example
of the chiral magnetic effect. An absolute stabilization
of metastable and unstable walls in a large magnetic field is found.
A possible higher dimensional generalization of the
chiral magnetic effect is mentioned.

\vspace*{.05cm}

\end{titlepage}

\section{Introduction}

Metastable brane-type extended objects such as  walls and strings are an
essential part of the nonperturbative spectrum in gauge theories, and are also
frequently discussed in a geometrical engineering of  SUSY gauge theories  and
in cosmology. The metastability of such extended defects is ensured by the well
known  semiclassical exponential suppression factor in their decay rate. A
somewhat less studied problem is that of the decay of these defects while being
populated by  additional degrees of freedom including the degrees of freedom
associated with defects of lower dimension. In this situation an interesting
question arises concerning the fate of such additional degrees of freedom  and
the effect of their presence on the decay rate.

In this paper we discuss three problems of such type: the decay of a metastable
axion wall in an external magnetic field, the decay of walls in dense matter,
and the decay of a nonabelian string with the excitation. It is known that due
to the  one-loop Goldstone-Wilczek current \cite{gw} an axion wall in an
external magnetic field develops a homogeneous density of the charged
fermions~\cite{sikivie,kogan,voloshin,it}. The question we  investigate here is
whether the fermions stay localized on the remaining part of the domain wall
during its decay or escape into the bulk? We find that in the leading
approximation the fermions leave the domain wall and provide a very clear
pattern of the chiral magnetic effect (CME) which currently attracts a considerable
attention related to the observed charge separation effect at RHIC \cite{khar}.
We shall thus argue that the decay of domain wall provides an effective chiral
chemical potential responsible for the chiral magnetic effect.

The second problem is related to the decay of mesonic walls in QCD. The
nontrivial density of the Skyrmions is generated in the external magnetic field
on  $\pi^0$  walls, and the field prevents the wall from nonperturbative
decay above some $B_{crit}$~\cite{ss}. There is a new point in the mesonic
wall decay in  dense matter. It is known \cite{sz} that in this case due to
the anomalous term in the action the current is generated along the axion-like
string at the boundary of the wall.  In the decay problem the axion string is the boundary of the hole and
therefore there is the circular boundary current during the decay. We analyze the
impact of the created current on the magnetic field.

The third problem in our discussion concerns the decay process of a nonabelian
string with excited internal $CP_N$ degrees of freedom. The
Lagrangian of the worldsheet theory supports a kink excitation \cite{sy}
which can exist
in isolation by itself in SUSY theories and as kink-antikink bound states  in a generic
non-SUSY case. We shall argue that in the nonabelian string decay the
kink excitations on the string decrease the decay rate.

The paper is organized as follows. In Section 2 we consider the process of the
axion domain wall decay .
The decays of mesonic walls in QCD are considered in Section 3.
In Section 4 we comment on the decay of nonabelian string  with the
emphasis on the role of kink excitations. The analogue of the CME in higher
dimensions is briefly discussed in Section 5.

\section{Decay of axion-like domain walls in $D=3+1$ theories}
In this Section we shall discuss the decay of the abelian
domain walls in the magnetic field.
Let us discuss the decay of axion  wall in the theory with the
Lagrangian for the axion field $a(x)$ given by
\beq
L= f_{a}^{-2}\, \left [ {1 \over 2} \, (\partial a)^{2} + m_a^2 \cos a(x) \right ]
\eeq
The model also contains charged fermions interacting with the
axion as
\beq
L_{f}= \bar{\psi}\left [ i(\partial_{\nu} - i\,  A_{\nu})\gamma^{\nu} - m_f \,
 e^{i \, a(x) \,\gamma^5}\right ] \psi
\eeq
with $A_\nu$ being the potential of the electromagnetic field. (The coupling $e$ is absorbed into the normalization of $A$.)
An integration over the fermionic field gives rise to the anomalous interaction between the axion and the electromagnetic field~\cite{gw} described by the Lagrangian

\beq
L_{anom}={1 \over 16 \pi^2} \, \epsilon_{\mu \nu \lambda \sigma} \, A_{\mu}F_{\lambda \sigma}\, \partial_{\nu} a
\eeq

A derivation of this term through the analysis of the fermionic modes
in the external field in the simplified model for the fermions
with the Lagrangian
\beq
L= \bar{\psi} \left[ i(\partial_{\nu} - i  A_{\nu})\gamma^{\nu} - \mu_1(z)
- i\mu_2(z)\gamma^5 \right ] \psi
\label{simple}
\eeq
can be found in \cite{voloshin}. The variation of $\mu_2$ breaks the
CP parity similarly to the axion field.

If the axion model admits a wall solution, the anomalous term
yields the density of the electric charge at the domain wall \cite{sikivie}.
The details of the domain wall solution are not important and the surface density of the
induced charge in the background magnetic field is equal to
\beq
q=\frac{B \Delta a}{4\pi^2}
\label{fcharge}
\eeq
where B is the magnetic field perpendicular to the axion wall
and $\Delta a$ is the total variation of the axion field across the wall. Therefore a constant external magnetic field creates a homogeneous
density of the induced electric charge on the wall. In the simplified
model the distribution of the induced charge density in the domain wall background equals to
\beq
\rho(z)=\frac{B}{4\pi^2}\, \frac{d}{dz}\arctan \frac{\mu_2(z)}{\mu_1(z)}
\eeq
For any realistic magnetic field its total flux through an infinite wall is zero, so that the total charge of such wall is zero as well. This behavior however is a result of exact cancelation between areas with positive and negative surface charges.

An axion wall is metastable and decays
through nucleation and subsequent expansion
of a hole bounded by an axion string. When considering this process in a
constant magnetic field, a natural questions arises concerning  the fate of the induced
electric charge during the decay as well as the back-reaction of the decay on the
background magnetic field. In order to analyze this issue we consider different components of the
GW current which are generated during the decay process. Let us assume for definiteness that
the hole is created at the origin in the $(x,y)$ plane

During the decay the axion field can be approximated as
\beq
a(z,r,t)= f(z)\, \theta(r-r(t))
\eeq
with some profile function $f(z)$, and  $r(t)$ is the time dependent radius of the expanding hole. Fermions are bounded by the domain wall
and therefore there is no reason for them to stay at the same point
when the hole is created. There are two logical possibilities:
the fermions fly away from the domain wall plane or are captured by the
axion string at the boundary of the hole. Using the GW expression we find that as the hole expands there arises a current perpendicular to the wall plane
\beq
J_z \propto (\partial_t a) \, B_z\propto f(z)\, \dot{r} \, \delta(r-r(t)) \, B_z
\eeq
which is clearly localized near the boundary of the hole.

The current $J_z$ is directed along the external magnetic field
and provides the explicit example of the chiral magnetic effect resulting
in the charge separation effect \cite{alekseev,khar}
which was recently a subject of intensive theoretical and experimental
studies. The domain wall decay process amounts to
the effective time dependent chiral chemical potential

$$\mu_{5,eff}=\partial_t a$$
The effective chiral chemical
potential is localized at the boundary of the hole - the axion string.
This is not a surprise since  the axion string in magnetic field is chiral,
i.e. there is an asymmetry between left and right modes of the fermions on the string.

We thus conclude that the fermions, initially localized at the wall, do fly away to the bulk in the decay process. One can perform a cross check of this conclusion by comparing the current `to the bulk' $J_z$ with the rate of the disappearance of the area of the wall that initially carried the fermions. Indeed, the rate at which the surface charge disappears through the growth of the hole is given by
\beq
dN = 2\pi \, q \,   r \, \dot{r} \, dt
\eeq
where $q$ is the surface charge density on the wall , $r(t)$
radius of the hole and the speed $\dot{r}$ is fixed by the bounce solution.
On the other hand if all fermions fly away  we must
have from the continuity equation
\beq
\frac{dN}{dt}= -\int d^3x \frac{dJ_{z}}{dz}
\eeq
Substituting the expression for the axion field corresponding
to the undeformed $O(4)$ symmetric bounce solution into the
anomalous current we obtain that the continuity equation is fulfilled.
This implies that all fermions fly away  from the wall plane
during the decay process
and there is no need for accumulating any fermions on the boundary of the hole.

It should be mentioned that in the considered case, where the area of the wall supporting the fermion charge changes and the fermions flow into the bulk, the charge conservation is implemented differently from the case of a fixed patch of the wall and varying magnetic field. In the latter situation the surface term of the GW current is not conserved by itself~\cite{callan} and its divergence
 is localized
at the string
\beq
\partial_{\mu}j_{\mu}\propto B \delta(r) ~.
\eeq
The apparent non-conservation of the current is  compensated by the accumulation of the charge on the axionic string at the boundary of the patch.

The escape of the fermions to the bulk produces an effect on the probability of the wall decay in a magnetic field. Indeed, the tunneling process is described by a spherical Euclidean bounce configuration which is determined from  the effective action
\beq
S_{eff}= 4\pi \, R^2 \, T_{string}- 4/3 \pi \, R^3 \, T_{wall}~,
\eeq
where R is the radius of the bounce. At the extremum of this action the surface of the bounce is described by
\beq
R_{crit}= \frac{2T_{string}}{T_{wall}}\, \qquad x^2 + y^2 + t_{E}^2 = R_{crit}^2
\eeq
with $t_E$ being the Euclidean time, so that the
coordinate $z$ transverse to the wall is not
essential in the bounce solution.

The transverse direction however becomes of relevance when the magnetic field is switched on.
In the leading approximation
the effect of the magnetic field can be taken into account as follows. While the fermions are
localized at the wall they are effectively massless and there is no energy associated with them. Once they escape
in the bulk each fermion costs energy equal to its mass $m_f$.
That is energetically the localization of the fermions
on the wall suppresses the decay probability.
From the (2+1) dimensional viewpoint the decay proceeds
with the energy loss since fermions escape from the wall. The fermion charge in a uniform magnetic field is proportional to the area of the wall (Eq.(\ref{fcharge})). Therefore the effect an energy loss due to emission of fermions can be described by replacing the wall tension by an effective one
\beq
T_{wall} \to T_{wall,eff}=T_{wall}- {B \over 2 \pi} \, m_f~,
\eeq
resulting in a suppression of the wall decay. Notably the decay is entirely suppressed, i.e. the wall is stabilized at the magnetic field exceeding the critical value
\beq
B_{crit}=2 \pi \, {T_{wall} \over m_f}~.
\eeq

Let us remark that for the case of an induced wall decay at a
non-vanishing energy the solution necessarily involves Minkowski
part in the time evolution. An example of such two-step
process  involving a resonance behavior
at  particular values of the energy
has been discussed in details in \cite{gv}. In the
current problem this would happen if in a process, e.g. in a collision, an excited state of axionic string is produced,
 which then tunnels through the Euclidean region and eventually reaches the classical expansion regime.

It can be noted also that in four dimensions it is possible to trade the pseudoscalar
field for a rank-two field $B_{\lambda \sigma}$ via the duality relation
\beq
\partial_\mu a = \epsilon_{\mu \nu \lambda \sigma}\,\partial_\nu B_{\lambda \sigma}~.
\eeq
In the presence of an external gauge field this relation gets modified
\beq
\partial_\mu a = \epsilon_{\mu \nu \lambda \sigma} \left ( \partial_\mu B_{\lambda \sigma}-A_\mu \, F_{{\lambda \sigma}}\right )~,
\eeq
where the second Chern-Simons term emerges at one loop. Then the
kinetic term of the axion at one loop yields the term
in the Lagrangian responsible for the GW current.
In terms of such dual description the domain wall implies a nontrivial
`electric' curvature of the rank-two field $H=dB$ localized along
the domain wall. The axionic string is charged with respect to
$B_{\lambda \sigma}$, that is in the dual representation the decay of the
axion domain wall corresponds to a Schwinger-type
production of axion strings in external field.


\section{Decays of  mesonic walls}
\subsection{Decay of $\pi^0$  domain walls}
In this Section we consider the decay of
walls in conventional QCD. One example of such object is provided by a wall built from $\pi^0$ mesons. A $\pi^0$ wall is not topological
and can be `unwound' inside the $SU(2)$ flavor
group. Furthermore such walls are absolutely unstable in the absence of the
magnetic field. However at $B>B_0= 3 m_{\pi}^2$ the wall becomes locally stable
and at $B>B_1= 16 \pi \, f_{\pi}^2 \, m_{\pi}/{m_N}$ a patch of such wall carrying a baryon number becomes the lowest energy state with baryon number~\cite{ss}.

The tension of the domain wall calculated at the explicit
solution~\cite{ss} reads as
\beq
T_{pwall}= 8 \, f_{\pi}^2 \, m_{\pi}~.
\eeq
A magnetic field $B$ applied perpendicularly to the wall generates a surface  density of the baryon charge
\beq
q_B = {B \over 2 \pi}~,
\eeq
which can be also viewed as a liquid of the Skyrmions on the surface~\cite{ss}.

The decay of the pionic wall implies a nontrivial baryonic
current
\beq
J_{\mu}= {1 \over 4 \pi^2} \, \epsilon_{\mu \nu \lambda \sigma} \partial_{\nu}\pi^0 F_{\lambda \sigma}
\eeq
flowing into the bulk similar to the electric current
in the  axion example. While escaping from the wall
the Skyrmions have mass $m_N$. Therefore the  effective
 wall tension can be found as
\beq
T_{eff}=T_{pwall} - q_B \,  m_N = 8 \, f_\pi^2 \, m_\pi - {B \over 2 \pi} \, m_N~.
\label{teff}
\eeq
One readily concludes from this expression that at $B > B_1$ the effective tension of the wall is negative, so that the decay is energetically impossible and the wall is absolutely stable.
It can be noted that a somewhat similar behavior has been  observed
for the decay of electric strings in magnetic
field~\cite{chernodub}.

\subsection {Wall decay in QCD at high density}

At high baryon density the ground state of QCD is in color-flavor
locking (CFL) phase and the system develops
color superconductivity
(see \cite{alford} for a review). The theory at large baryon
chemical potential
$\mu$ is in the weak coupling regime and the
dynamics of the low-energy
degrees of freedom can be calculated perturbatively.
In particular, the existence of a $\phi$ domain wall
can be justified \cite{ssz} from
the effective Lagrangian for the Goldstone mode $\phi$
of $U(1)_A$ symmetry which is spontaneously broken by the condensate in the
color-superconducting vacuum state.

The explicit form of the Lagrangian in two-flavor case reads as follows
\beq
L_{dense}= f^2 \, \left [ (\partial_0 \phi)^2 - u^2 (\partial_i \phi)^2 \right ]
- a \, \mu^2 \, \Delta^2 \, \cos \phi
\eeq
where $a$ is dimensionless and vanishes in the limit $\mu\rightarrow \infty$, and $u$ is the speed of sound: $u^2=1/3$.
The parameters of the Lagrangian are
\beq
f^2=\frac{\mu^2}{8 \pi \, u^2}~,
\eeq
and $\Delta$ is the value of the gap. The tension of the wall can be derived
immediately from the effective Lagrangian
\beq
T_{wall}= 8 \, \sqrt{2a} \, u \, f \, \mu \, \Delta~.
\eeq
In the CFL phase the potential term in the Lagrangian
of the lightest meson gets modified as
\beq
V_{CFL} = -  \tilde{a} (\frac{m_s}{\mu}) \, \mu^2 \, \Delta^2 \, \cos \phi~,
\eeq
where $m_s$ is the mass of the strange quark. Thus the tension
of the wall in the CFL phase acquires  additional
$m_s$ dependent factor.

The decay of the domain
walls in the dense QCD matter  has some peculiarities.
In particular, one can notice that in the dense matter there is an anomalous
Chern-Simons term in the  Lagrangian of the pseudoscalar meson
proportional to the chemical potential $\mu$
\cite{sz,ss}:
\beq
\delta L = \frac{e}{24 \pi^2} \, \mu \, \epsilon_{0 \nu \lambda \sigma} \, \partial_\nu \phi \, F_{\lambda \sigma}~.
\eeq
An immediate consequence of this term is that there is an
electric current circulating along the axial string \cite{sz,met}
\beq
J=\frac{\mu e}{12\pi}~.
\label{current}
\eeq
It can be noted that the current does not depend on the value of external field. This current can be derived
by the summation over the fermion modes \cite{met} similar
to the calculation of the induced charge in the magnetic field \cite{voloshin}.

The main difference from the wall decay discussed in the previous Section is that in dense matter
there necessarily is a current along the hole boundary identified with the
axial $\phi$ string. The current plays a two-fold role. Firstly, its existence implies
that during the decay process not all of the  fermions populating
the wall  fly away from the plane. Some of them are captured
by the axial string at the boundary instead. Secondly, the circular current
induces  magnetic field inside the hole and the direction of the field depends on the
sign of the chemical potential.
One thus concludes that
in dense matter magnetic effects in the decay
of the wall are necessarily essential since the field is generated
by the induced current circulating along the  hole boundary.
The tunnelings starts at zero energy, so that there is no cusp at the
bottom of the bounce configuration. However contrary
to the axion and $\pi^0$-mesonic walls in the vacuum, the current along the boundary makes
it impossible to describe the  outflow of the fermion energy
by an appropriately modified effective wall tension as in Eq.(\ref{teff}).

One more issue is related to the angular momentum conservation
during the domain wall decay. It was argued in \cite{sz}
that the domain wall in dense matter carries an
induced constant angular momentum per unit area
\beq
{\cal M}=\frac{\mu}{6\pi}~,
\eeq
so that it may appear that during the decay the angular momentum of the part of the wall turning into a hole is `lost'. However,
the current along the hole boundary also yields an induced
angular momentum and can in fact ensure the conservation.

\section{Nonabelian string decay}
The strings are quite common objects corresponding to effective solutions
to the equations of motion in various models. The problem of their
decays can be formulated, and a detailed analysis of the decay of an ANO string in the Abelian
Higgs model can be found in \cite{Shifman:2002yi}. As of yet
only the decay of such ANO Abelian effective strings  has been
considered. However more general  stringy solutions exist both in the SUSY and non-SUSY
theories in the color-flavor locking phase.
Their key feature is the existence of additional degrees of freedom
due to the nontrivial embedding of the nonabelian string into the gauge group, which
amounts to the orientational moduli providing $CP_N$ degrees of freedom on the
worldsheet \cite{sy}.
Thus the problem that parallels the discussion in the previous section is that of the
fate of the $CP_N$ degrees of freedom living on the nonabelian string during
the decay. There is however an essential difference between nonabelian strings in SUSY and non-SUSY cases.

The worldsheet theory is built from an $N$-component
complex field $n^i$ subject to the constraint
\beq
n_i^*\, n^i =1\,.
\label{lambdaco}
\eeq
The Lagrangian has the form
\beq
{ L} = \frac{2}{g^2}\, \left[
\left(\partial_\mu - i A_\mu\right) n^*_i
\left(\partial_\mu + i A_\mu\right) n^i
-\lambda \left( n^*_i n^i-1\right)
\right]\,,
\label{one}
\eeq
where  $\lambda$ is the Lagrange multiplier
enforcing the condition (\ref{lambdaco}).
At the quantum level this constraint is effectively eliminated
and $\lambda$ becomes dynamical. Moreover,
$A_\mu$ is an auxiliary field which at the classical level enters the Lagrangian
with no kinetic term. A  kinetic term is generated, however,
at the quantum level, so that the field $A_\mu$
becomes dynamical too.

One can also add to the Lagrangian a $\theta$ term of the form
\beq
{ L}_\theta = \frac{\theta}{2\pi }\, \varepsilon_{\mu\nu} \partial^\mu A^\nu
=\frac{\theta}{2\pi }\, \varepsilon_{\mu\nu} \partial^\mu \left(
n_i^*\partial^\nu n^i
\right)\,.
\label{two}
\eeq

In the SUSY case the worldsheet theory is supplemented by fermionic terms.
There are  (N-1) vacuum states in the model and, respectively,  kinks interpolating between those states.
In a sense the bounce configuration
provides the decay of the $CP_N$ model on a plane into a pair of  $CP_N$
models on semiplane. If the string is in one of the ground states, the decay
corresponds to creation of a pair of heavy monopole and antimonopole
located at the ends of two semi infinite strings.

We consider here the decay of a string in the presence of kinks. We start with discussing the case where only one kink is present
 and question whether this kink induces the string
decay or suppresses it instead. Clearly, the kink, being
a monopole in the Higgs phase, cannot escape into the bulk, contrary to
the previously considered situations, since the kink does not exist in the bulk at all.
Hence it can be only at the end of one semi infinite string   or be located somewhere
on the string worldsheet. The location at the end is impossible since
it would yield an additional nonabelian magnetic flux in the emerging hole
which cannot be the case due to the Meissner effect. One therefore
concludes that the kink has to be outside the decay region and
that it thus effectively suppresses the decay.

In non-SUSY D=4 gauge theories
there is only a single  vacuum state, so that a single kink solution is impossible. However the
spectrum involves a kink-antikink pair, which corresponds, from
the four dimensional viewpoint,
to a monopole-antimonopole pair localized at the nonabelian string \cite{gsy}.
Such pair can not exist in the bulk theory as well, so that in similarity
to the SUSY case the string excitations suppress the decay rate.
In the non-SUSY case one can also introduce the $\theta$-term in the bulk
theory, which makes its way to the worldsheet theory of the nonabelian string as well~\cite{gsy}. In the worldsheet theory the $\theta$ term induces a
constant abelian electric field of the auxiliary gauge field $A(x)$ along  string.
In the string decay the electric field is completely
screened in the emerging hole. One therefore concludes that the dyons have to be created
at the ends of the string in this case.

Note that the explicit calculation of the decay probability
would require a detailed
knowledge of the quantum boundary $CP_N$ model.

\section{An analogue of CME in higher dimensions}

In this Section we shall make some remarks concerning
possible higher dimensional generalization of the chiral
magnetic effect. To this aim let us consider a
3 brane providing a four dimensional worldvolume. One could think
of a brane world setup where the 3 brane is embedded
in higher dimensional, e.g. $D=5$, space-time with additional
two- form field. The problem we would like to comment on
is the decay of the 3 brane worldvolume due to creation
of the  hole surrounded by a domain wall.

Let us consider the specific situation where the 3 brane
is embedded into the $D=5$ space-time with the transverse coordinate
$z$. The question we address here is that of the anomalous
flow from the $3+1$ brane worldvolume into the bulk
during the decay process. In the two-form field background
there is the current
\beq
J_{\mu} = \epsilon_{\mu \nu \lambda \rho \sigma} \, \partial_{\nu}\phi  \, H_{\lambda \rho \sigma}
\eeq
emerging from the anomalous one-loop term in the lagrangian
\beq
\delta L = \int d^5 x \, \phi \, F_{\mu\nu}*H_{\mu\nu}~,
\eeq
where $\phi$ is a pseudoscalar field similar to the axion
and $H=dB$ is the curvature of the 2-form field.

In a constant curvature $H$ a homogeneous electric charge
is generated in the 3d space volume with the density
\beq
\rho \propto \epsilon_{zijk} \, H_{ilk}~,
\eeq
where it is assumed that the 3 brane interpolates
between different values of the field $\phi$
similar to an axion wall.
One thus may wonder what happens to the charge
during the decay. The corresponding current
providing the flow from the brane reads as
\beq
J_z\propto \partial_0 \phi  \epsilon_{zijk}H_{ilk}
\eeq
For a $3+1$ dimensional observer the charge
seemingly is not conserved. However having in mind
a holographical interpretation of the fifth
coordinate one could interpret this flow as an RG one.

\section{Summary}
In this note we have discussed  decay of  defects
in external field. It was shown that in a magnetic
field the axion domain wall evaporates all
induced  electric charge into the bulk. Such decay
of the axion wall provides an explicit example
of the chiral magnetic effect where the axion strings
are responsible for the chiral chemical potential.

The decay of the mesonic walls in  magnetic field
has some peculiarities.  Namely, the decay probability of the pionic walls
is  suppressed by a magnetic field and
above a critical value of the field the wall is
non-perturbatively stable. In the CFL phase at high
density the current along the boundary
of the hole in the $\eta$-meson walls is generated
decreasing the initial magnetic field.
An example of  similar phenomena in a higher-dimensional
case has been discussed. This potentially can be
interesting for  holographic description of
gauge theories.

Our consideration here was limited to the leading approximation in the
electromagnetic coupling constant. It would be
interesting  to take into account the emerging currents on the
field distribution around the hole in the next order in the electromagnetic interaction. t
One more potentially interesting effect is related to possible evaporation
of  closed axion or mesonic strings from the decaying wall
rather than the evaporation of particles. This subleading effect
can potentially be of relevance in an analysis of stability
of mesonic walls.

The processes discussed in this note can be of interest 
for astrophysical applications. In particular the unstable 
cosmic domain
walls could be stabilized in the space regions with large magnetic
fields.

\section*{Acknowledgments}
A.G. thanks FTPI where the part of the work has been done for  hospitality and support.
The work of A.G. is supported in part by the grants PICS- 07-0292165,
RFBR-09-02-00308 and
CRDF -  RUP2-2961-MO-09. The work of M.B.V. is supported in part by the DOE grant DE-FG02-94ER40823.

\end{document}